\documentclass{article}
\usepackage[T1]{fontenc}

\usepackage{amsfonts}
\usepackage{graphicx}

\newtheorem{defi}{Definition}[section]
\newtheorem{theo}{Theorem}[section]
\newtheorem{stw}{Lemma}[section]
\newtheorem{conj}{Conjecture}[section]

\textwidth\paperwidth
\advance\textwidth -55mm
\advance\oddsidemargin 30mm
\advance\evensidemargin 25mm
\advance\topmargin 2cm
\title{On total communication complexity of collapsing protocols for pointer jumping problem}
\date{May 28, 2014}
\author{Micha{\l} Jastrz\k{e}bski}

\begin{document}
\maketitle

\begin{abstract}
This paper focuses on bounding the total communication complexity of collapsing protocols for multiparty pointer jumping problem ($MPJ_k^n$). Brody and Chakrabati in \cite{bc08} proved that in such setting one of the players must communicate at least $n - 0.5\log{n}$ bits. Liang in \cite{liang} has shown protocol matching this lower bound on maximum complexity. His protocol, however, was behaving worse than the trivial one in terms of total complexity (number of bits sent by all players). He conjectured that achieving total complexity better then the trivial one is impossible. In this paper we prove this conjecture. Namely, we show that for a collapsing protocol for $MPJ_k^n$, the total communication complexity is at least $n-2$ which closes the gap between lower and upper bound for total complexity of $MPJ_k^n$ in collapsing setting.
\end{abstract}


\section{Introduction}

Communication complexity is a tool which has been proven successful in proving lower bounds in wide range of areas. Since its inception in \cite{yao} (where it was used as a measure of amount of communication which has to be performed between two separate groups of processors in order to compute certain functions) it has been a field of constant research with an increase of scientific interest in the recent years. Being defined as a very general problem of multiple parties trying to collaboratively compute an outcome of a function using as small amount of inter-communication as possible, the notion was being used for examining properties of bounds on complexity of streaming algorithms (consult \cite{chakr08, guha07}), circuit complexity (for instance \cite{karchmer98}), lower bounds on data structure performance (such as \cite{miltersen}) and much more. Great, short introduction to communication complexity can be found in \cite{ab} whereas a definite treatment of the topic together with many application examples on application is contained in \cite{nisan96}. \\

\noindent In the general multiparty setting we have players $PLR_1,PLR_2,...,PLR_k$ sharing the input $(x_1, x_2,.., x_k)$. The goal is to compute value $f(x_1,...,x_k)$ for a particular function $f$, with the \emph{complexity} being defined as a minimal number of bits they have to communicate among themselves to do so. Two most common models according to which information is shared between players are
\begin{itemize}
\item \textbf{Number-In-Hand} ($NIH$) in which player $i$ knows only $x_i$
\item \textbf{Number-On-the-Forehead} ($NOF$) where player $i$ knows all $x_j$ for $j \neq i$ ($x_i$ is written on his forehead).
\end{itemize} 

\noindent We focus on the $NOF$ model, which was introduced by Chandra, Furst and Lipton in \cite{chandra83}. One of the reasons for conducting research on $NOF$ multiparty model is the fact that obtaining a \emph{strong} communication complexity lower bound for a particular function $f$ would cause its non-membership in complexity class \textbf{$ACC^0$} (informally - functions computed by Boolean cicruits of constant depth and polynomial size where set of possible gates include the \emph{modulo counter} for some fixed constant).
Details of this exposition can be found for instance in \cite{ab}. \\

\noindent Pointer-Jumping problem is considered as a good candidate for this task. In this case inputs $x_1,...,x_k$ are essentially functions and players want to collaborate in computing the result of convoluting them. We will describe the problem more precisely together with the current state of research in following sections. It is worth noting that in the $NOF$ model players share \emph{\textbf{almost}} all the information contained in the input. This means that it should be very hard to prove lower bounds in such model. Another reason for considering the Pointer-Jumping problem is its generality - one may see the address shifting, multiplication and many others as a special case of this function.

\noindent 

\section {Pointer Jumping Problem}

In this work we concentrate on the classical, boolean version of pointer jumping problem ($MPJ_k^n$). The problem is build on top of the graph $G_n^k$ being a directed graph with $k + 1$ layers of vertices. The first layer, 0, consists of just one vertex $v$. Each of the consecutive $k-1$ layers (numbered $1,2,.., k-1$) contains exactly $n$ vertices. The last layer consists of two vertices labeled $0$ and $1$. These labels denote the possible outcome of the problem. In graph $G_n^k$ each vertex from layer $i$ is connectected with a directed edge with each vertex from layer $i+1$.

The input to $MPJ_k^n$ problem is a graph $K_n^k$ being a subgraph of $G_n^k$ (on the same set of vertices) where each vertex apart from layer $k$ has \emph{outdegree} $1$. Outcome, or the solution of the problem, is the label of the vertex in the layer $k$ (which is $0$ or $1$) which is obtained by following the directed path starting from vertex $v$ (layer $0$) through all the layers of $K_n^k$. 

Consider $k$ players numbered $PLR_1,PLR_2,..,PLR_k$ sitting in a circle and having a blackboard. For each $i$, player $PLR_i$ has the set of edges comming from layer $i-1$ to $i$ written on his forehead. This means that each of players knows all the edges of the graph apart from those assigned to him. Computation in this model is being held in a following way : players communicate in a fixed order $PLR_1, PLR_2,..., PLR_k$. Each of the players, based on the information communicated by previous players, as well as information he can observe (data writted on foreheads of all other players) outputs some bits and writes them on the blackboard. The last player, namely $PLR_k$ outputs only one bit $0$ or $1$ being the outcome of his computation. The outcome is supposed to be the result of pointer jumping process on the graph $K_n^k$. Figure 1.1 shows the sample instantiation of Pointer-Jumping problem with a graph $K_4^4$. We have $4$ players, each of them stores edges marked by respective dotted rectangles. Following black dots we find out that $PLR_4$ should output $1$ as the outcome of computation.

\begin{figure}
\centering
\includegraphics[scale=1.0]{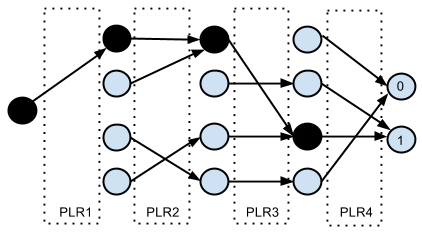}
\caption{Sample instatiation of the $MPJ_4^4$ problem}
\end{figure}

When defining a playing protocol $P$, and denoting strings of bits outputed by consecutive players by $p_1,..,p_{k-1}$ we may define a \emph{total communication cost} of protocol $P$ as 
$$C_{total}(P) = \sum_{1\leq i < k}|p_i|$$
the total number of  bits output by all players. We may be also interested in \emph{maximal communication cost} which is 
$$ C_{max}(P) = \max_{1\leq i < k}(|p_i|)$$

We are mainly interested in communication complexity of a problem (and not a particular protocol), which we may define as
$$C_{total}(MPJ_k^n) = \min_{P}(C_{total}(P))$$
I analogous was we define the maximal communication complexity. We consider only deterministic protocols, namely we require the protocol $P$ to be always correct when computing the solution. \\

\noindent \textbf{Remark}. The simplest protocol for solving $MPJ_k^n$ has total cost $C_{total} = n$. Indeed, the player $PLR_{k-1}$ just outputs all the values written on the forehead of $PLR_k$ (which is a string of n-bits). Now, player $PLR_k$ can follow the pointer up to layer $k-1$, and then apply information output by $PLR_{k-1}$ to compute the final answer. This shows that $C_{total}(MPJ_k^n) = O(n).$ \\

\noindent \textbf{Remark}. It is worth noticing that the order in which players speak out is crucial. In any other case one may come up with a protocol of total cost $O(\log n)$. Indeed, if the order differs from $PLR_1,..., PLR_k$ then there exist indices $i < j$ such that $j$ speaks before $i$. The protocol involves only players $j$ and $i$:
\begin{enumerate}
\item $PLR_j$ : follows the pointers up to layer $j-1$. Outputs the index of the resulting vertex, which takes $log(n)$ bits
\item $PLR_i$ : as $i < j$ he can just follow the pointers from the vertex outputted by $PLR_j$ up to the last layer
\end{enumerate}

\section {Previous results}

Let's start with a simple case $k=2$. It is easy to show that $C_{total}(MPJ_2^n) = n$. Indeed, otherwise one can easily construct to different pairs of inputs $(x_1, x_2)$ and $(x_1, x'_2)$ such that $f(x_1, x_2) \neq f(x_1, x'_2)$ but both for $x_2$ as well as for $x'_2$ $PLR_1$ outputs the same value. $PLR_2$ seeing in both cases exacltly the same data has to output the same final value of $f$ which is a contradiction. Problem gets much more complex for $k \geq 3$. \\

\noindent As disscussed in the previous section, proving a \emph{strong} lower bounds on communication complexity for $MPJ_k^n$ would result in proving $MPJ_k^n \notin ACC^0$. Precisely speaking we have the following
\begin{theo}
If there exist constants $\alpha, \beta > 0$ such that for multiparty total communication complexity in $NOF$ model for \textbf{$k = n^{\alpha}$} we have $C(MPJ_k^n) = \Omega(n^\beta)$ then $MPJ_k^n \notin ACC^0$. 
\end{theo}

\noindent For more details please consult \cite{nisan96} or \cite{bt}. As for now such lower bound is not known. In fact, the best known lower bound for $MPJ_k^n$ comes from Viola and Widgerson \cite{viola07} and states
\begin{theo}
\label{theo:lower}
$C_{total}(MPJ_k^n) = \Omega(n^{1/(k-1)} / k^{O(k)})$.  In particular $C_{total}(MPJ_3^n) = \Omega(\sqrt{n})$.
\end{theo}

\noindent Note that this result, even though it is the best currently known lower bound, provides non-trivial result only for small values of $k$, in particular for $k = n^\alpha$ the resulting bound is already trivial. On the other side for a long time time following conjecture was open 
\begin{conj}
There exists nondecreasing function $\alpha:\mathbb{Z^+} \to \mathbb{R^+}$ such that for all  $k$, $C_{total}(MPJ_k^n) = \Omega(n / \alpha(k))$
\end{conj}

\noindent Looking at the hard lower bound for $k = 2$ as well as some weaker versions of communication model (where such statement is true) one could claim that the conjecture should hold also for a full $NOF$ model. However, 
Pudlak, Rodl and Sgall in \cite{pudlak97} by considering a problem $PERM_k$ (which is a subcase of $MPJ_k^n$ where functions given to players $PLR_2,...,PLR_{k-1}$ are all permutations) showed a tricky protocol proving a sublinear bound

\begin{theo}
\label{theo:perm}
$C_{total}(PERM_3^n) = O \left( n \frac{\log \log n}{\log n}\right)$
\end{theo}

\noindent This result has been extended by Brody and Chakrabarti in \cite{bc08} to the full $MPJ_k^n$ problem:

\begin{theo}
\label{theo:sublinear}
$C_{total}(MPJ_k^n) = O\left( n(\frac{\log \log n}{\log n})^{(k-2) / (k-1)}\right)$ and in particular
$$ C_{total}(MPJ_3^n) = O(n \sqrt{\log n / \log \log n}) $$
\end{theo}

\noindent The theorem clearly falsifies the \emph{linear lower bound} conjecture. These are the first and by far best upper bounds for the $MPJ_k^n$ problem. Proving better complexity bounds of pointer jumping and narrowing the current gap seems to be a deep and difficult problem. Thus, different simplified versions of $MPJ_k^n$ are actively being considered in hope for much stronger bounds and as motivating examples for research of original problem:
\begin{itemize}
\item Restricting the set of functions which can be provided as an input for the player
\item Restricing the $NOF$ model of communication
\end{itemize}
We describe such kind of modifications in following sections.

\section {Note on variants of $MPJ_k^n$}

\noindent First simplified variant has been already described in the previous section and involves restricting functions owned by players $PLR_2,..,PLR_{k-1}$ to permutations. In this variant, a sublinear protocol was given (Theorem \ref{theo:perm}) which motivated Brody and Chakrabati to develop a sublinear protocol for general $MPJ_3^n$ problem (and uses the $PERM_3^n$ protocol as a black box) \\

\noindent Another simplified problem is $TPJ_k$ which imposes a tree structure on the graph $G_k^n$.  In this setting, the underlying graph is a $k$-height tree where each vertex from layer $0$ through $k-2$ has $n^{1 / (k-1)}$ children laying in the next layer and vertices from layer $k-1$ have two children, namely $0$ and $1$. One can easily see, that such tree is in fact a subgraph of regular $G_k^n$ used in a general $MPJ_k$ model. Indeed, every layer contains not more than $n$ vertices (layer $k-2$ contains exactly $n$ vertices). Interestingly, the lower bound coming from the theorem is actually a lower bound for $TPJ_k$, so that Viola and Widgerson show Theorem \ref{theo:lower} in an even stronger setting
\begin{theo}
$C_{total}(TPJ_k) = \Omega(n^{1/(k-1)} / k^{O(k)})$
\end{theo}

\noindent Remarkably, such a lower bound for $TPJ_k$ is also the best known lower bound for $MPJ_k^n$ which suggest that it can be significantly improved. On the other hand, due to the fact of such big restrictions imposed on structure of functions, one may easily show that
\begin{stw}
$C_{total}(TPJ_k) = O(n^{1/(k-1)})$
\end{stw}
\noindent which implies that the bounds are tight in this particular case. Indeed, it is enough for $PLR_2$ to output the result of following the pointers from every vertex of layer $1$. As there are only $n^{1/(k-1)}$ vertices in this layer, such number of bits is enough for player $PLR_k$ to correctly produce the answer. \\

\noindent Another version of $MPJ_k^n$ being considered is $\widehat{MPJ_k^n}$, a version of $MPJ_k^n$ in which the last layer consists of $n$ vertices (and not of $2$ as in the case of $MPJ_k^n$) so that the output is a number $1, 2,.., n$ rather than binary. For this version, the trivial protocol has complexity $n \log(n)$. Damn,  Jukna and Sgall in \cite{damn98} proposed a simple protocol with complexity $O(n \log^{(k)}n)$ for $\widehat{MPJ_k^n}$ where $\log^{(k)}$ means the $k$-th iterated logarithm. Unfortunately their protocols shows no nontrivial bounds for $MPJ_k$.

\section {Restricted protocols}
Due to the difficulty of improving lower and upper bounds for general $NOF$ models, different restricted models with restrictions are being considered. Restricting the communication model makes the bound range easier to prove and provides additional techniques and ideas potentially leading to succesfull attacks to general $NOF$ model.
We discuss three different models in this section together with respective results. \\

\noindent \textbf{Conservative protocol} \\
\noindent Here we require that a player $PLR_i$ can see all the layers ahead of him but only a \emph{composition} of layers behind him. To be precise, if by $f_j$ we denote a edges from layer $j-1$ to $j$ (which is in fact a function) then player $PLR_i$ can see all the functions $f_{i+1},f_{i+2},...$ but \textbf{only} a composition of previous functions, namely $f_{i-1} \circ f_{i-2} \circ ... \circ f_{1}$. Intuitively, $PLR_i$ knows the result of following the pointers of all previous layers up to level $i-1$, but cannot see the way this result has been obtained. \\

\noindent Conservative protocols were introduces by Damn, Jukna and Sgall in \cite{damn98} while noticing that the protocol for $\widehat{MPJ_k^n}$ is in fact conservative. Restricting protocols in such way, one can prove much stronger bounds on complexity, as shown in \cite{damn98}:
\begin{theo}
In a conservative model $C_{total}(MPJ_k^n) = \Omega(n/k^2)$
\end{theo}
\noindent which shows that such restrictions lead to almost linear lower bounds. \\

\noindent \textbf{Myopic protocol} \\
Protocol designed by Damn, Jukna and Sgall possess another important property. $PLR_i$ can see only edges from layer $i$ to $i+1$. Such protocol is called \emph{myopic} due to the described property. Formally, using the functional notation described above, $PLR_i$ can produce the ouput  based on (apart from the information coming from previous players) functions $f_1,.., f_{i-1}, f_{i+1}$. The model has been introduced in \cite{gron06}. J. Brody in \cite{brody09} proved following theorem

\begin{theo}
In conservative setting for $MPJ_k^n$ some player must communicate $n/2$ bits and there exists protocol achieving this bound. Moreover, players have to communicate at least $n$ bits in total. In other words $C_{total}(MPJ_k^n) = n$ and $C_{max}(MPJ_k^n) = n/2$.
\end{theo}

\noindent \textbf{Collapsing protocol} \\
\noindent Another very interesting restriction imposed on the protocol is essentialy reverting the \emph{conservative} model. This time we allow player $PLR_i$ to know all the previous layers, but restrict his view of the layers in front of him to its composition. Namely, $PLR_i$ sees $f_1,..,f_{i-1}$ as well as $f_k \circ ... \circ f_{i+1}$. Such restriction seems to be particularly interesting mostly due to the sublinear protocol from Theorem \ref{theo:sublinear}. In fact, deeper look at the protocol (exposition of which can be found in Appendix) shows that during communication, \emph{every} player apart from $PLR_1$ behaves in a \emph{collapsing} way - he sees only the composition of layers in front of him. \\

\noindent The following question seems natural : \emph{Is the fact that one of players is noncollapsing a necessary condition for obtaining sublinear upper bound?}. Brody and Chakrabati in \cite{bc08} prove that, indeed, this is the case. We have the following

\begin{theo}
In a collapsing protocol for $MPJ_k^n$ there exists a player communicating at least $n - \frac{1}{2}\log_2n-2$ bits. In other words $C_{max}(MPJ_k^n) \geq n - \frac{1}{2}\log_2n-2$.
\end{theo}

\noindent Thorem shows that even one noncollapsing protocol makes a fundamental difference in complexity. Obtained lower bound is still lower than obvious upper bound $n$. The gap has been closed by Liang in \cite{liang} where the following is shown

\begin{theo}
\label{theo:coll}
There exists a collapsing protocol for $MPJ_k^n$ where each player communicates at most $n-\frac{1}{2}\log_2n +1$ bits which matches the lower bound up to an additive constant.
\end{theo}

\noindent Liang's protocol doesn't however say anything about the bound on \emph{total} complexity of the problem. In fact, total complexity of Liang's protocol is $n + \frac{1}{2} \log_2 n$ which is worse then obvious upper bound of $n$. 
Theorem \ref{theo:coll} shows us of course $$n - \frac{1}{2} \log_2 n \leq C_{total}(MPJ_k^n) \leq n .$$
Liang in his paper posed an open question : \emph{Is the lower bound on \textbf{total} communication complexity achievable?} \\

\noindent In this work we provide a proof providing \textbf{negative} answer to this question. We will show that a trivial protocol of complexity $n$ is the best we can achieve in terms of total communication complexity.

\section {Our contribution}

Main contribution of this paper is resolving the open problem stated in Liang's paper \cite{liang}. We show the following theorem
\begin{theo}
\label{theo:main}
For a \textbf{collapsing} protocol the total communication complexity of $MPJ_k^n$ is bounded by 
$$C_{total}(MPJ_k^n) \geq n -2.$$
\end{theo}

\noindent The proven lower bound shows that there are no collapsing protocols achieving better total complexity (up to a small additive constant) than the trivial one. This way we close the previously known gap ($n - \frac{1}{2} \log_2 n \leq C_{total}(MPJ_k^n) \leq n$) between upper and lower bound of total communication complexity of $MPJ_k^n$ in a collapsing model. \\ 

\noindent This section is devoted to proving Theorem \ref{theo:main}. We will use some techniques developed in $\cite{bc08}$, when proving the lower maximum complexity bound in collapsing setting.

\subsection {Preliminaries and formal notation}
We will formally define $MPJ_k^n$ problem here. By $[n]$ we denote set $\{1,2,...,n\}$  and for a string $x$, by $x^{(i)}$ we denote its i-th element. Formally the input of $MPJ_k^n$ problem is a tuple $(i, f_2, f_3,...,f_{k-1}, x)$, where $i \in [n]$, $f_i \in [n]^{[n]}$ for $2 \leq i \leq k-1$ and $x \in \{0,1\}^n$.  Note that $x$ can be seen both as a function $[n] \to \{0,1\}$ (which allows us to use functional notation) and as a $n$-element string of $0,1$ (thus we may write $x^{(i)}$). Formally 

$$ MPJ_k^n : [n] \times \left( [n]^{[n]} \right) ^{k-2} \times \{0,1\}^n \to \{0, 1\}$$

\noindent where $MPJ_k^n$ itself can be defined recursively in a formal manner
\begin {eqnarray}
MPJ_2^n(i,x) &=& x^{(i)} \nonumber \\
MPJ_k^n(i,f_2,f_3,...,f_{k-1},x) &=& MPJ_{k-1}^n(f_2(i),f_2,...,f_{k-1},x) \nonumber
\end{eqnarray}
In other words, $MPJ_k^n(i,f_2,..,x) = x \circ f_{k-1} \circ ... \circ f_2(i)$. Intuitively, we have already described what a playing protocol is what does it mean that a protocol is \emph{collapsing}. Here we describe it in a purely formal way.
A protocol will be called collapsing it there exist functions $P_1,...P_k$ (rules according to which players are producing outputs) 
\begin{eqnarray}
P_1:\{0,1\}^n \to \{0,1\}^{t_1}  \nonumber \\
\forall_{1 < i < k}P_i : \{0,1\}^{t_1} \times ... \times \{0,1\}^{t_{i-1}} \times [n] \times \left( [n]^{[n]} \right) ^{i-2} \times \{0,1\}^n \to \{0,1\}^{t_i} \nonumber
\end{eqnarray}
such that for any input $(s,f_2,...,f_{k-1}, x)$ we can define
\begin{eqnarray}
\alpha_1 &=& P_1(x \circ f_{k-1} \circ ... \circ f_2) \nonumber \\
\alpha_i &=& P_k(\alpha_1, ..., \alpha_{i-1}, s, f_2,...,f_{i-1}, x \circ f_{k-1} \circ ... \circ f_2) \nonumber
\end{eqnarray}
such that $\alpha_k = MPJ_k^n(s,f_2,...,f_{k-1}, x)$.

\noindent In such setting $\alpha_i$ are messages outputed by consecutive players and $C_{total}(P) = \sum_{1 \leq i \leq k-1}t_i$

\subsection {Proof of Theorem \ref{theo:main}}

\noindent We will prove our theorem by contradiction. Namely, after assuming that players send out in total less than $n-2$ bits, we will construct a \emph{fooling set}, pair of inputs $(i, f_2,...,f_{k-1}, x)$ and $(i, f_2,...,f_{k-1}, x')$ which are indistinguishable for all the players (all of them will have to output the same bits in both cases) but the result of $MPJ_k^n$ functions on those inputs don't match.

\noindent To start with, we need lemmas showing that for a function outputing only a certain amount of bits there must exist pair of elements (having some property) that the function cannot distinguish.

\begin {defi}
Let $x,y \in \{0,1\}^n$. We will say that $x < y$ iff for every index $1 \leq j \leq n$ we have $x^{(j)} \leq y^{(j)}$ and there exists index $j$ for which $x^{(j)}< y^{(j)}$. 
\end{defi}

\begin{stw} \label{stw:chain}Let $f: \{0,1\}^n \to [n]$. (One may also look at it as a function $f$ outputing $k < log(n+1)$ bits of information). Then there exists $x,y \in \{0,1\}^n$ such that $x < y$ and $f(x) = f(y)$.
\end{stw}

\noindent \textbf{Proof.} Suppose on the contrary that we have a function $f:\{0,1\}^n \to [n]$ such that for every $x < y$ we have $x \neq y$. Let us consider the following chain of elements. $x_0,x_1,....,x_n\in \{0,1\}^n$ such that $x_i$ is a string consisting of $i$-zeroes followed by $(n-i)$-ones. Formally, $x_i^{(k)} = 0$ for $k \leq i$ and $x_i^{(k)}=1$ for $k > i$.
It is easy to notice that for each $i < j$ we have also $x_i < x_j$ (it is a proper chain in the meaning of partial order created by "$<$"). This means that our function (based on the assumption) has to take different values for different elements of the chain. As the sequence consists of $n+1$ elements, the function has to take at least $n+1$ different values. Contradiction.

$\hfill \Box$

\noindent \textbf{Remark.} It is worth noticing that for $k = log(n+1)$ the previous statement does not hold. Indeed, consider a function counting number of ones in a string
$$f(x) = \sum_{1 \leq i \leq n}x^{(i)}.$$
Clearly, this function outputs $log(n+1)$ bits of information (as the values range from $0$ to $n$), but for each $x < y$ we have $f(x) < f(y)$. \\

\begin {defi}
\cite{bc08} For strings $x,y \in \{0,1\}^n$ and $a,b$, define the sets
$$  I_{ab}(x, y) = \{j \in [n] : (x^{(j)}, x^{(j)}) = (a,b)\} $$
A pair of strings $(x, y)$ is \textbf{crossing pair} if for all $a,b \in \{0,1\}$, $I_{ab}(x, y) \neq \emptyset$
\end {defi}

\begin{stw} \label{stw:crossing} (Proof in \cite{bc08})
Let $f:\{0,1\}^n \to \{0,1\}^t$ where $t \leq n - 0.5\log n - 2$. Then there exists a crossing pair of elements $x,$ such that $f(x) = f(y)$.
\end{stw}

\noindent Lemmas \ref{stw:chain} and \ref{stw:crossing} show that when the output of function $f$ is small enough, then one can find pair of elements $x, y$ such that $x < y$  (or respectively $(x,y)$ - crossing) such that $f(x) = f(y)$, which means that $f$ cannot distinguish them. Let us also use a following definition simplifying notation : 

\begin{defi}\cite{bc08}
A string $x \in \{0,1\}^n$ is said to be consistent with $(f_1,...,f_j, \alpha_1,...,\alpha_j)$ if in prototol $P$, for all $h \leq j$, $PLR_h$ sends the message $\alpha_h$ on seeing input $(i=f_1,...,f_{h-1}, x \circ f_j \circ ... \circ f_{h+1})$ and previous messages $\alpha_1,...,\alpha_{h-1}$.
\end{defi}

\noindent Our proof is using the idea of finding a \emph{fooling pair} of strings, which we define as

\begin{defi}
Pair of strings $(x,y)$, $x,y \in \{0,1\}^n$ we will call a $j$-fooling pair, if there exists $(f_1,...,f_j, \alpha_1,...,\alpha_j)$ such that both $x$ and $y$ are $(f_1,...,f_j, \alpha_1,...,\alpha_j)$-consistent and 
$x \circ f_j \circ ... \circ f_1 \neq y \circ f_j \circ ... \circ f_1$.
\end{defi}

\noindent Of course, proving existence of $(k-1)$-fooling pair shows that $P$ is not a valid protocol. Indeed, in that case the player $PLR_k$ will see exactly the same information for both $x$, and $y$, thus he has to produce the same answer, which contradicts the definition of $(x,y)$. We will now show a series of statements allowing for iterative construction of \emph{fooling pairs}.

\begin{stw}
\label{stw:push}
For a given protocol $P$, if there is a $(x,y)$, $j$-fooling pair, such that $x < y$ then if $PLR_{j+1}$ sends less than $n-2$ bits, then there exists a $(x_1, y_1)$, (j+1)-fooling pair. 
\end{stw}

\noindent \textbf{Proof.}  Let us take $(f_1,...,f_j, \alpha_1,...,\alpha_j)$ that $(x,y)$ is consistent with. We will now construct $(x_1, y_1)$ and $f_{j+1}, \alpha_{j+1}$ such that $(x_1, y_1)$ is $(j+1)$-fooling and is consistent with 
$(f_1,...,f_j,f_{j+1}, \alpha_1,...,\alpha_j, \alpha_{j+1})$. Player $PLR_{j+1}$ (after seeing all the previous messages and information) sends less than $n-2$ bits, this means that there exist strings $x_1, y_1 \in \{0,1\}^n$ such that 
\begin {eqnarray}
x_1^{(1)} &=& y_1^{(1)} = 0 \nonumber \\
x_1^{(2)} &=& y_1^{(2)} = 1 \nonumber \\
x_1 &\neq& y_1 \nonumber
\end{eqnarray} 
\noindent and $x_1, y_1$ are indistinguishable by bits output by $PLR_{j+1}$. Last condition means that there exists index $i > 2$ such that $x_1^{(i)} = 0$ and $y_1^{(i)} = 1$ (in the opposite case we just swap $x_1,y_1$). We now want to construct mapping $f_{j+1}$ satisfying $x \circ f_{i+1} = x_1$ and $y \circ f_{i+1} = y_1$. Our assumption $x < y$ shows that $I_{01}(x,y) \neq \emptyset$ whereas $I_{10}(x,y) = \emptyset$. Thus we may define our function
\begin {eqnarray}
f_{j+1}(s) = 1, s \in I_{00}(x,y) \nonumber \\
f_{j+1}(s) = 2, s \in I_{11}(x,y) \nonumber \\
f_{j+1}(s) = i, s \in I_{01}(x,y) \nonumber
\end{eqnarray}
\noindent One may easily notice that $f_{j+1}$ satisfies our conditions and setting $\alpha_{j+1}$ as an output of player $PLR_{j+1}$ upon his information, we obtained a pair $(x_1, y_1)$ which is $(j+1)$-consistent. Indeed, choosing $x_1, y_1$ to be undistinguishable by $PLR_{j+1}$ based on his information, we guarantee what $\alpha_{j+1}$ is the same for both of those strings.

$\hfill \Box$

\begin{stw}
\label{stw:crosspush}
For a given protocol $P$, if there is a $(x,y)$, $j$-fooling pair, then if $PLR_{j+1}$ sends less than $n-0.5 \log n -2$ bits, then there exists a $(x_1, y_1)$, (j+1)-fooling pair such that $(x_1, y_1)-crossing$.
\end{stw}

\noindent \textbf {Proof.}  Schema of the proof is similar to the one in Lemma \ref{stw:push}. We are interestied in constructing $f_{j+1}, \alpha_{j+1}$ as well as a crossing pair $x_1,y_1 \in \{0,1\}^n$ such that $(x_1,y_1)$ is consistent with $(f_1,...,f_j,f_{j+1}, \alpha_1,...,\alpha_j, \alpha_{j+1})$. As we know, $PLR_{j+1}$ sends (after seeing all other information) less than $n - 0.5 \log n - 2$ bits which means that  (according to Lemma \ref{stw:crossing}) one can construct a crossing pair $(x_1, y_1)$ which is undistinguishable by bits output by $PLR_{j+1}$. Such property gives us nonemptiness of sets $I_{00}(x_1,y_1),I_{01}(x_1,y_1),I_{10}(x_1,y_1),I_{11}(x_1,y_1)$. Thus, after choosing a representant from each of those sets (respectively $i_{00},i_{01},i_{10},i_{11} $) we may define $f_{j+1}$ in a following way
\begin {eqnarray}
f_{j+1}(s) = i_{00}, s \in I_{00}(x,y) \nonumber \\
f_{j+1}(s) = i_{11}, s \in I_{11}(x,y) \nonumber \\
f_{j+1}(s) = i_{01}, s \in I_{01}(x,y) \nonumber \\
f_{j+1}(s) = i_{10}, s \in I_{10}(x,y) \nonumber 
\end{eqnarray}
\noindent This way, again we have constructed a valid function $f_{j+1}$ for which $x \circ f_{i+1} = x_1$ and $y \circ f_{i+1}  = y_1$ and the way we have chosen $x_1,y_1$ guarantees that $\alpha_{j+1}$ match in both cases.

$\hfill \Box$

\noindent In a way analogous to previous lemma we may prove the following

\begin{stw}
\label{stw:chainpush}
For a given protocol $P$, if there is a $(x,y)$, $j$-fooling pair and $x<y$, then if $PLR_{j+1}$ sends less than $\log (n+1) -2$ bits, then there exists a $(x_1, y_1)$, (j+1)-fooling pair such that $x_1<y_1$.
\end{stw}

\noindent \textbf {Proof.} Analogous to proofs of Lemmas \ref{stw:push} and \ref{stw:crosspush}

$\hfill \Box$

\noindent \textbf{Proof of Theorem \ref{theo:main}} \\

\noindent Let us assume existence of protocol $P$ of cost $C_{total}(P) < n-2$. We will consider two cases. \\

\noindent \textbf {Case1:  $|PLR_1| \geq log(n+1) - 2$ }. Of course we have also $|PLR_1| < n$ (due to our assumption).  This means that there exist a $(x,y)$, 1-fooling pair. Indeed  It is enough to take two strings $x \neq y$ on which $PLR_1$ outputs the same value. Such pair must exist as $PLR_1$ outputs less than $n$ bits. $x \neq y$ gives us the existence of index $j$ such that $x_j \neq y_j$. It is now enough to take $f_1 := j$. $\alpha_1$ is the message produced by $PLR_1$, which (due to our construction) is the same both for $x$ and $y$.  If each of players $PLR_2,..,PLR_{k-1}$ sends less than $n- 0.5 \log n - 2$ bits, then using Lemma \ref{stw:crosspush}  multiple times we are able to consecutively generate pairs $(x, y)$ $j-fooling$ for $j = 1, 2, ..., k -1$, ending up with a $(x', y')$, $(k-1)$ - fooling pair, which contradicts validity of the protocol. \\

\noindent Thus, there must exist player $PLR_i$ for $2 \leq i \leq k-1$ which outputs more than $n - 0.5 \log n -2$ bits. Together with player $PLR_1$, this gives us already total complexity $ > n$. \\

\noindent \textbf {Case2:  $|PLR_1| < log(n+1) - 2$ }. This time, we can use Lemma \ref{stw:chain} to begin our construction. Indeed, from Lemma \ref{stw:chain} we obtain a pair of strings $x < y$ on which $PLR_1$ produces the same output. Now setting $f_1 = j$ such that $j \in I_{01}(x,y)$, and $\alpha_1$ as the message produced by $PLR_1$, we obtain a $(x,y)$, $1$-fooling pair. \\

\noindent Based on our assumption on total cost of protol, we will prove inductively the following statement \\

\noindent \textbf {Statement} \emph{For every $1<i \leq k-2$, $PLR_i$ outputs less than $\log(n+1)  - 2$ bits and there exists a $(x_i,y_i)$, $i$-fooling pair for which $x_i < y_i$}.  \\

\noindent We know that the statement is true for $i = 1$. Knowing that it is true for all $i \leq j$ we will prove it for $j+1$ (where $j+1 \leq k-2$). Assume that player $PLR_j$ outputs less than $\log n -2$ bits and there is a $(x_j, y_j)$, $j$-fooling pair such that $x_j < y_j$. Our assumption on total complexity gives us that $PLR_{j+1}$ outputs less than $n - 2$ bits. Using Lemma \ref{stw:push} we receive a $(x'_{j+1}, y'_{j+1})$, $(j+1)$ - fooling pair.  If we now assume that each of players $PLR_{j+2},...,PLR_{k-1}$ outputs less than $n - 0.5 \log n - 2$ bits, then using Lemma \ref{stw:crosspush} multiple times, we will be able to construct pairs $(x'_{j+2}, y'_{j+2}),...(x'_{k-1}, y'_{k-1})$ which are respectively $j_2, j+3,..., k-1$-fooling. This would contradict the validity of the protocol (nonexistence of $(k-1)$-fooling pairs). This means that at least one of $PLR_{j+2},...,PLR_{k-1}$ outputs at least $n - 0.5 \log n - 2$. This, on the other hand, implies that $PLR_{j+1}$ outputs less than $\log n - 2$ bits (according to our assumption on total complexity). Using Lemma \ref{stw:chainpush} we obtain also a $(x_{j+1}, y_{j+1})$, $(j+1)$-fooling pair for which $x_{j+1} < y_{j+1}$ and thus prove our \emph{Statement}. \\

\noindent The \emph {statement} shows us that $PLR_{k-2}$ outputs less than $\log n - 2$ bits and there exists a $(x_{k-2}, y_{k-2})$, $(k-2)$-fooling pair such that $x_{k-2} < y_{k-2}$. Due to the total complexity assumptions, player $PLR_{k-1}$ has to output less than $n-2$ bits. This, however, according to Lemma \ref{stw:push} implies existence of $(x_{k-1}, y_{k-1})$, $(k-1)$-fooling pair which, as we know, contradicts validity of protocol and gives a contradiction proving the main theorem.

$\hfill \Box$

\section {Conclusions and further work}

We have proven that there are no collapsing protocols achieving a total communication complexity for $MPJ_k^n$ (in $NOF$ model) better than a trivial one (up to an additive constant). This closes the gap between lower bound of $n - 0.5 \log n -2$ and the upped bound and at the same time answers the open problem posed in work of Liang, \cite{liang}. \\

\noindent Main open problems remain showing that $MPJ_k^n \notin ACC^0$ through finding lower bounds with high number of players ($k = n^\alpha$). Another challenging problem is tightening the lower and upper bounds in the general $MPJ_k^n$ problem where the the gap between $(\sqrt{n})$ versus $(n\sqrt{\log\log n / \log n})$ is still very big. \\

\noindent We believe that showing tight bounds for protocols with additional restrictions can push the development of general protocols for $MPJ_k^n$ further allowing for resolving two very challenging problems in this area.

\end{document}